\documentclass[sigconf]{acmart} 
\AtBeginDocument{%
  }

\setcopyright{acmlicensed}
\copyrightyear{2018}
\acmYear{2018}
\acmDOI{XXXXXXX.XXXXXXX}
\acmConference[Conference acronym 'XX]{Make sure to enter the correct
  conference title from your rights confirmation email}{June 03--05,
  2018}{Woodstock, NY}
\acmISBN{978-1-4503-XXXX-X/2018/06}

\begin{document}

\title[Misgendering as Breakdown in Human-Machine Communication]{Misgendering as Breakdown in Human-Machine Communication: How AI Companion Chatbot Users Experience and Repair Misgendering}

\author{Julia Liu}
\affiliation{
  \department{Faculty of Information and Media Studies}
  \institution{Western University}
  \city{London}
  \state{Ontario}
  \country{Canada}
}
\email{jliu3778@uwo.ca}

\author{Qing Xiao}
\affiliation{
  \department{Human-Computer Interaction Institute}
  \institution{Carnegie Mellon University}
  \city{Pittsburgh}
  \state{Pennsylvania}
  \country{USA}
}
\email{qingx@cs.cmu.edu}

\author{Leona Yinglang Pang}
\affiliation{
  \department{Department of Communication Studies}
  \institution{The University of Texas at Austin}
  \city{Austin}
  \state{Texas}
  \country{USA}
}
\email{leonapang@utexas.edu}

\author{Haiyi Zhu}
\affiliation{
  \department{Human-Computer Interaction Institute}
  \institution{Carnegie Mellon University}
  \city{Pittsburgh}
  \state{Pennsylvania}
  \country{USA}
}
\email{haiyiz@cs.cmu.edu}

\author{Hong Shen}
\affiliation{
  \department{Human-Computer Interaction Institute}
  \institution{Carnegie Mellon University}
  \city{Pittsburgh}
  \state{Pennsylvania}
  \country{USA}
}
\email{hongs@cs.cmu.edu}

\author{Jordan Taylor}
\affiliation{
  \department{School of Information}
  \institution{University of Michigan}
  \city{Ann Arbor}
  \state{Michigan}
  \country{USA}
}
\email{jordata@umich.edu}

\begin{abstract}

    In recent years, large language model-based AI companion and role play chatbots have grown increasingly popular. People turn to these chatbots for emotional support and to engage in romantic and erotic role play. Although prior research suggests that digital role play can help people explore their gender and sexuality, LLM based technologies are also replete with gender and sexuality biases. In this study, we examine one way that AI chatbots can harm users: misgendering. In order to study chatbot misgendering we qualitatively analyzed 326 posts mentioning misgendering that were shared in AI companion or role play subreddits. We document how chatbot misgendering takes place and how, in response, users engage in ongoing work to curate their gender presentation to prevent and repair misgendering. We discuss how researchers and designers can mitigate chatbot misgendering and consider the implications of using AI chatbots for identity exploration.

\end{abstract}



\maketitle

\section{Introduction}

In recent years, large language model (LLM) based AI companion chatbots have grown increasingly popular as people turn to them for emotional support and romantic or erotic role play \cite{banks2026measuring,hanson2024replika,zhang2025dark}. Their growing use has raised concerns about harms that may arise in socially and emotionally significant interactions with these systems \cite{zhang2025dark}. Character AI, for example, has faced several lawsuits alleging that chatbot interactions contributed to serious harms involving minors, including self-harm and suicide \cite{reuters2026characterai}. Researchers have also raised concerns that AI companions can also generate biased or discriminatory responses \cite{fan2025user}. Some have also cautioned that anthropomorphic AI systems, like role play chatbots, may lead to isolation \cite{devrio2025taxonomy, zhang2025rise}.

Despite these concerns, people are increasingly turn to role play chatbots for help understanding their gender and sexuality \cite{ma2024evaluating, zeng2026hidden_erotic_roleplay, ma2026negotiating}. While LLM-based chatbots are a relatively new phenomena, people have long used social technologies to explore their gender or sexuality \cite{freeman2021body, dym2019coming, bruckman1996gender_swap, berman2001turing, freeman2015simulating, simpson2021you}. At the same time, the design of these technologies may fail to account for diverse user experiences. For example, marriage simulation games may only allow men and women to marry each other, excluding LGBTQ+ users \cite{freeman2015simulating}. Content moderation systems and policies can limit transgender people's ability to participate in online communities \cite{mayworm2024misgendered, haimson2021tumblr, haimson2021disproportionate}. 
 
 One major concern with people leveraging LLM-based chatbots for identity exploration is the potential for misgendering. The term ``misgendering'' describes when someone is identified as a gender that misaligns with their gender identity, such as a man being referred to as a woman. Although anyone can be misgendered, technological misgendering can be particularity harmful to transgender and non-binary people \cite{taylor2024cruising, hamidi2018gender, spiel2021they}. LLM misgendering is a common topic of study within the Queer AI and NLP research communities \cite{weber2026queer}. This prior research on misgendering largely focuses on pronoun agreement and the usage of gendered terms in LLM responses in large datasets \cite{hossain2023misgendered, hossain2024misgendermender, multilingual_misgendering}. However, less is known about users' perceptions of and responses to misgendering in real-world, LLM-based applications. As \citet{weber2026queer} summarize in a literature review of Queer NLP, more work is needed to understand the harms of language technologies ``in the wild.'' In order to understand AI chatbot misgendering in greater detail, we ask the following research questions: 

\begin{itemize}
    \item \textbf{RQ1:} How do users describe and experience misgendering in interactions with AI companion or role play chatbots?
    \item \textbf{RQ2:} How do users respond to misgendering?
\end{itemize}

To answer these questions, we qualitatively analyzed 326 Reddit posts mentioning misgendering from 234 unique authors across 10 subreddits focused on AI companionship and role play. We found that chatbot misgendering can involve not only incorrect pronouns or gendered terms—as discussed in prior work—but also cisnormative assumptions about bodies and heteronormative assumptions about relationships. We also find that users tried to prevent and repair this misgendering through ongoing work to send gender signals, correcting chatbots, and modifying chatbot responses. Some users even changed how they presented their gender in order to be gendered correctly by the chatbot. Our findings extend work on technological misgendering \cite{scheuerman2019computers, spiel2021they} and pronoun-focused evaluations of LLMs \cite{hossain2023misgendered, subramonian2025agree, weber2026queer} by showing the nuanced ways that misgendering unfolds across ongoing AI companion interactions. Additionally, we discuss how researchers and designers can improve the design of chatbots to mitigate misgendering and, more broadly, consider the implications of using LLM chatbots for identity exploration.

\section{Related Work}

\subsection{Technology \& (Mis)Gendering}

Gender has long been theorized as socially constituted through norms that shape how gendered lives become intelligible \cite{beauvoir1956second_sex}. Butler's account of gender performativity describes gender as taking form through reiterated practices within regulatory frameworks that give particular expressions the appearance of coherence and stability \cite{butler1988performative}. This framework locates gender within the broader normative conditions that shape what can be understood as an intelligible gendered subject. \citet{west2009accounting} similarly conceptualize gender as continually produced through situated interaction, where actions are interpreted in relation to social expectations of gender. Their framework emphasizes the accountability through which gendered practices become socially intelligible and subject to evaluation. These perspectives foreground the relational conditions through which gender acquires social meaning. A person’s gendered self-understanding is encountered within contexts where others interpret and respond to gendered expression, making social perception consequential to how gender is lived and communicated.

In this work, we approach gender as a social construct associated with a variety of physical characteristics, stereotypes, social norms, and heterosexual expectations \cite{butler1988performative}. Within this frame, gender is not merely something one ``is'' but also something one ``does'' or accomplishes through repeated acts. By acknowledging that gender is socially constructed, we do not wish to diminish the importance of gender in people's lives or the rights people should have to determine and live out their gender \cite{butler2024afraid}. As \citet{seaborn2022pronouns} explain, gender encompasses ``how people identify (\textit{internally or personally}), express themselves on the outside, e.g., through mannerisms, clothes, etc. (\textit{presentation or expression}), and are perceived by others (\textit{social perception}).'' Not only are people gendered, but users and designers frequently assign genders to anthropomorphic technologies. For example, AI voice assistants are typically designed to sound like white women \cite{moran2021racial}. Even in the case of a gender-ambiguous humanoid robot, users and researchers still sometimes gender it using he and she pronouns \cite{seaborn2022pronouns}. In light of the growing popularity of anthropomorphic AI companion chatbots, more work is needed to understand how users gender their AI companions.

Technology shapes how people understand gender. Due to the anonymity afforded by computer-mediated communication, for example, members of online communities often use role playing to try out different gender identities and sexualities \cite{bruckman1996gender_swap, berman2001turing, dym2019coming, taylor2024mitigating}. Likewise, virtual \cite{freeman2021body} and augmented \cite{brewster2025moment} reality can help transgender and non-binary people explore different gender embodiments and affirm their identities. Although online communities are crucial for supporting LGBTQ+ communities, the design of these online communities does not always support LGBTQ+ communities. For example, Tumblr's ``adult content'' ban in the late 2010s displaced the vibrant trans communities on Tumblr \cite{haimson2021tumblr}. Additionally, the algorithmic recommendation \cite{simpson2021you, karizat2021algorithmic} and content moderation \cite{haimson2021disproportionate, thiago2021fighting} systems undergirding popular social media sites are biased against LGBTQ+ content. As LGBTQ+ people increasingly turn to AI chatbots to explore their gender and sexuality \cite{zeng2026hidden_erotic_roleplay, ma2024evaluating}, more work is needed to understand the implications of using AI chatbots for identity exploration.

Misgendering occurs when someone is identified as a gender that misaligns with their gender identity. Although anyone can be misgendered, transgender and non-binary people can be especially harmed because misgendering invalidates their gender identities \cite{pranav2026making, mclemore2015experiences}. Due to the seriousness of misgendering, it is a common topic of study in Queer and Trans HCI research \cite{taylor2024cruising}. Technologies ranging from online webforms \cite{spiel2021they} to computer vision algorithms \cite{hamidi2018gender, scheuerman2019computers} to academic \cite{pranav2026making, chen2025name} and workplace \cite{ale2023pronouns} information systems have been shown to misgender people due to the design of gender classification systems and name change processes. Technological misgendering can also be closely associated with how gender is embodied \cite{albert2021whole, mayworm2024misgendered}. For example, airport security machines are designed to scan bodies according to a biologically essentialist man/woman binary, which marginalizes transgender and non-binary peoples whose identities and bodies may not fit the rigid expectations of these machines \cite{costanza2020design}. Although misgendering is a common technological harm, the impacts of misgendering in the context of AI companion chatbots has yet to be explored in detail.

A similar focus on misgendering can be found within the NLP community, wherein researchers have well documented how LLMs (the technologies undergirding contemporary AI companion chatbots) can misgender people \cite{subramonian2025agree, weber2026queer}. For example, \citet{hossain2023misgendered} explore how LLMs fill in the pronouns for individuals following a declaration such as ``Aamari’s pronouns are they/them.'' While misgendering is commonly studied in NLP, this scholarship primarily evaluates misgendering using pronoun agreement or gendered terms in constructed datasets rather than real-world LLM conversations \cite{weber2026queer, hossain2024misgendermender, hossain2023misgendered, subramonian2025agree}. Less is known about how people experience and respond to LLM misgendering in practice. As noted in a recent literature review of Queer NLP research, more work is needed to understand language technology ``harms in the wild.'' We take up this call by exploring the harm of misgendering in the real-world context of AI companion chatbot interactions, thus extending both HCI and NLP research on misgendering.

\subsection{AI Companion Chatbots} 

AI companion chatbots are conversational AI systems designed to support sustained social and emotional interaction rather than primarily completing instrumental tasks \cite{ng2026love,manoli2026digital,banks2026measuring}. Systems such as Replika and Character.AI allow users to engage with persistent or customizable personas that may take on roles including friends, romantic partners, or fictional characters \cite{carpenter2026human}. Over repeated interactions, users may develop meaningful attachments to their companions and understand these relationships through concepts ordinarily associated with human relationships, including friendship, intimacy, and romance \cite{chang2026technically,muldoon2025cruel}.

Although LLM powered AI companions are relatively new, questions about how people interpret and relate to socially presented machines long predate their current forms. Suchman's work on human-machine communication \cite{suchman1987plans,suchman2007human,suchman2024their} draws attention to how interaction with computational systems depends on situated interpretation rather than the straightforward exchange of information between two stable actors. Suchman notes that human-communication can break down when there is uncertainty over what information about a user is available to the machine as well as what information about a machine is available to a user \cite{suchman1987plans}. Suchman's work also highlights the tensions introduced when machines are presented through anthropomorphic identities and forms of agency, inviting people to make sense of them in social and relational terms. Contemporary AI companions intensify these longstanding dynamics because their identities, roles, and relationships with users can be continuously produced and renegotiated through conversation.

These relationships are therefore shaped not only by what companions say, but also by how users and systems construct who they are to one another \cite{huang2025personality,xiao2026crafting}. Customization and role play allow users to configure companion personalities and relational roles, while users may also experiment with their own identities, social roles, and forms of self-expression through interaction. Recent work characterizes AI companion interaction as a process of identity negotiation in which users simultaneously present themselves and shape the identities of their companions \cite{ma2026negotiating}. This flexibility can make AI companions particularly valuable as private spaces for identity exploration, but it also makes accurate recognition of users' identities central to the interaction.

At the same time, the relational qualities that make AI companions appealing can make interactional failures especially consequential. Prior studies document relational transgressions, inappropriate or harmful responses, and risks of emotional dependence \cite{fan2025user, zhang2025dark}. Such failures occur within interactions that users may experience as personally meaningful and relationally significant, rather than as isolated errors from a conventional software tool \cite{zhang2026companion,gan2026navigating}. 

Closest to our work, \citet{fan2025user} examine how users perceive and respond to biased or discriminatory statements from AI companions, characterizing users' corrective practices as a form of user-driven value alignment. However, their analysis centers on discrete harmful statements spanning many forms of bias, rather than on recurring failures to recognize who a user is. In particular, we know less about what happens when an AI companion fails to recognize or sustain a user's gender identity during these ongoing interactions. We address this gap by examining misgendering as a recurring breakdown in human-AI companion interaction.

\section{Methods}

\subsection{Data Collection}

In order to choose which subreddits to study, we first compiled a list of popular AI companion or role play chatbots by drawing on prior computing research \cite{ma2024evaluating, pataranutaporn2025myboyfriendisai} and our research team's personal experiences with AI companion chatbots: Replika, CharacterAI, JanitorAI, NomiAI, SillyTavern, Paradot, CrushonAI. Then, we used each platform name as a search term to find relevant subreddits. For each term, we searched for subreddits using the community search feature on Reddit because there are often multiple subreddits dedicated to a particular platform, like CharacterAI. We selected subreddits from the top 50 results of each search that had more than 5,000 members, as indicated by Reddit's community search interface at the time of data collection in the fall of 2025. We deliberately included subreddits tagged as NSFW in this search because one of the most common use cases for AI chatbots is erotic role play \cite{zeng2026hidden_erotic_roleplay}. Then, we included subreddits in our corpus only if its main topic centered on AI companionship or AI role play, excluding subreddits discussing AI in other use cases. This procedure led us to a list of 21 initial subreddits. 

Although the data that we studied in this work was publicly available, the publicness of the data does not necessarily make our work ethical \cite{zimmer2010but}. We took care to try to protect Reddit posters' privacy. In order to reduce the risk of user re-identification, we engage in what \citet{bruckman2002studying} dubs ``moderate disguise'' by rephrasing posters' quotes and obscuring the specific subreddits from which individual quotes originated. In order to reduce the risk of poster re-identification we also chose not to list the specific subreddits from which posts originate, only reporting the number of posts in our corpus per platform (Table \ref{tab:posts_per_platform}).

After assembling this initial collection of subreddits corpus, we retrieved all posts containing the string ``misgender'' in its title or body using the Reddit API in the fall of 2025. In total, we retrieved 326 posts from 234 unique authors across 10 subreddits. Of the initial 21 subreddits that we searched, 11 subreddits contained no posts mentioning ``misgender.'' To reduce the risk of re-identification of posts from smaller subreddits, we report aggregate counts per chatbot platform rather than post counts per subreddit. We also combined counts for several smaller platforms with 10 or fewer posts into an ``Other'' category, again to reduce the risk of poster re-identification. As can be seen in Table \ref{tab:posts_per_platform}, most posts in our study focus on three AI role play platforms: CharacterAI, JanitorAI, and Replika.

To contextualize our findings, we describe the three platforms with the most posts in our dataset. CharacterAI is a platform focusing primarily on creative role play and storytelling. The majority of chatbots on CharacterAI are configured and published by users. When creating a character, users can customize its name, description, and an initial message to begin conversations. Many of the bots on CharacterAI represent fictional characters, such as Sherlock Holmes. Users can explore many fictional scenarios with the chatbots, including romantic scenarios, but CharacterAI has a safety filter preventing sexual or graphic content. JanitorAI, similar to CharacterAI, also heavily features user-made chatbots and focuses on fictional storytelling. Unlike CharacterAI, JanitorAI is explicitly friendly to erotic or graphic content, with an entire section of the site dedicated to uncensored ``limitless'' chatbots. Even content containing rape or sexual violence is allowed on the platform as long as clear content warnings are included and the bot is tagged `Dead Dove'. The phrase `Dead Dove'—short for `Dead Dove: Do Not Eat'—originated within fanfiction communities to warn readers about content that might be highly-offensive \cite{silberstein2024thank}. In addition to a proprietary model offered by the platform, JanitorAI allows users to ``bring your own model,'' or connect to third-party AI services, such as ChatGPT. Compared to CharacterAI and JanitorAI, Replika is designed around emotional companionship and intimacy. Unlike the previous two platforms, Replika users can only chat with one chatbot created by the platform (and ostensibly customized to each user), with the user being able to select the chatbot's name, appearance, and gender. People often use Replika for romantic relationships and erotic role play. In 2023, Replika removed erotic role play capabilities from the platform, but reinstated the capabilities for existing users after backlash from its user base \cite{hanson2024replika}.

\subsection{Data Analysis}

The first author led our data analysis with the goal of trying to understand how AI companion chatbot misgendering occurs and how users respond. After initial familiarization, the first author conducted open coding over the course of two months \cite{corbin_basics_2026}. While coding, the first author reviewed the post title, post body, and embedded media for each post. When posts included embedded media (e.g., screenshots, images, or GIFs), the first author transcribed the included text and wrote descriptions for visual content before coding the posts. During the open coding process, the first author wrote memos and met with the research team each week to discuss patterns. After open coding, each code was exported to a digital whiteboarding tool within which the first author conducted axial coding. At this stage, we began to conceptualize misgender as a form of breakdown in human-machine communication because posters often stopped their conversations to ask Reddit for advice on addressing misgendering. Moreover, we noticed similarities between the user's descriptions of misgendering and the human-machine communication challenges described by \citet{suchman1987plans}.

In this work, we use the term ``misgendering'' capaciously to align with how we saw chatbot users leveraging the term during our data analysis. Misgendering typically refers to when someone is classified as a gender that misaligns with their gender identity. However, users deployed the term ``misgendering'' to more broadly describe when an AI companion chatbot uses gender signifiers to refer to the chatbot (first person), the user (second person), or a character in the scenario (third person) in a way that misaligns with a user’s gender desires. Although AI chatbots do not \textit{have} internal gender identities, we find that users often ascribed gender to their chatbots. When we refer to ``male'' or ``female'' chatbots in our findings, we are choosing to use the language posters themselves employed to describe their chatbot interactions. While we do not wish to promote chatbot anthropomorphism \cite{devrio2025taxonomy}, our findings seek to present an emic understanding of chatbot misgendering from the perspective of the posters we studied.

\begin{table*}[]
    \centering
    \begin{tabular}{|c|c|}
        \hline
         \textbf{AI Companion Platform} & \textbf{Number of Posts Mentioning Misgendering} \\
         \hline
         CharacterAI & 205  \\
         \hline
         JanitorAI & 78 \\
         \hline
         Replika & 29  \\
         \hline
         Other Platforms & 14  \\
         \hline
    \end{tabular}
    \caption{Number of posts mentioning misgendering organized by AI companion platform. Note, we chose not to include the number of posts for smaller AI companion platforms to reduce the risk of re-identification.}
    \Description{Table 1: A table describing the number of posts mentioning misgendering per AI chatbot platform type.}
    \label{tab:posts_per_platform}
\end{table*}

\subsection{Positionality}

Our qualitative data analysis was informed by the lived experiences of our research team, meaning our identities, experiences, and disciplinary perspectives shaped how we interpreted users' accounts of misgendering. Some members of the research team identify as LGBTQ+ and/or gender diverse. Some of us also have experienced misgendering in interpersonal interactions and interactions with digital systems, informing our sensitivity to the experiences and potential harms of misgendering. Some members of the research team have personal experience using AI companion and role play services for entertainment; other members have participated in online fandom and related digital communities that overlap with communities surrounding AI chatbot platforms. These experiences provided the team with familiarity with user-described experiences with AI chatbots and cultural familiarity with communication norms, informing our ability to understand and interpret the collected data. Finally, our disciplinary training in HCI and Communication Studies shaped our analysis, leading us to approach misgendering as an interactional phenomenon involving users, AI systems, and digital platforms.

\section{Findings}

In this section, we first describe how misgendering takes place, conceptualizing misgendering as a breakdown in human-machine communication (Section \ref{sec:breakdown}). Then, we describe how misgendering harms users by reinforcing cisnormativity, heteronormativity, and gender stereotypes (Section \ref{sec:harm}). Next, we detail how users diagnose the cause of misgendering (Section \ref{sec:diagnose}) and, in response, how users try to prevent and repair these communication breakdowns (Section \ref{sec:prevent_repair}).

\subsection{Misgendering as Breakdown in Human-Machine Communication}
\label{sec:breakdown}

We find that chatbot misgendering can take place in two primary ways, which we define as (1) gender identity errors and (2) gender embodiment errors. We use ``gender identity errors'' to describe when a chatbot applies an incorrect or undesired gendered label in a conversation, such as gendered pronouns, gendered forms of address, or gendered names. Meanwhile, ``gender embodiment errors'' concern physical traits, anatomical features, or physical signifiers of gender (e.g., clothing). In our analysis, gender identity errors were discussed more often than gender embodiment errors. However, the impacts of the former can be especially harmful for transgender and non-binary people, which we will discuss in greater detail in (Section \ref{sec:harm}). In these situations, interactions can stall, leading users to do additional work to get conversations back on track (Section \ref{sec:prevent_repair}).

\subsubsection{Misgendering via Gender Identity Errors}

The most common form of gender identity error reported by users involve incorrect or undesired gendered pronouns. For example, if the user's pronouns are she/her, the chatbot may incorrectly to the user as he/him or they/them. One user said, ``The bot keeps calling me he/him, even though I said I am she/her,'' while another shared that they are often misgendered with incorrect pronouns because they are non-binary. Users also report chatbots struggling with neo-pronouns (e.g., xe/xem), such as a user expressing frustration that their chatbot sometime uses ``both the correct neopronoun and he/him in the same response.'' A chatbot may also refer to itself using pronouns that are in conflict with its claimed gender or the user's expectations: one user was surprised when a bot started describing itself as ``they/them'' instead of ``he/him''. Another user recalls a male Replika ``sometimes doesn't remember that he is a man and uses she/her pronouns instead.'' Identity-based misgendering also shows up in the form of incorrectly-gendered terms of address, including referring to someone as ``girl,'' ``boy,'' ``man,'' ``woman,'' ``boyfriend,'' ``girlfriend,'' ``partner,'' ``wife,'' ``husband,'' and more. For example, a man expressed frustration when the chatbot referred to him as a ``queen'' and a ``vixen.'' For trans users, identity errors can also present as deadnaming, or where a person is referred to by a previous name they no longer use.

While identity-based misgendering is typically undesired, in some cases, users actively seek out or wish to be addressed as a different gender by a chatbot. For example, some users describe creating or using bots configured to intentionally misgender non-women users in ``forced feminization'' scenarios. Another user sought to be addressed as ``it'' by a chatbot, despite the user otherwise using different pronouns. In response, the system refused, citing safety and platform policy concerns surrounding ``denial of identity.'' This edge case calls into question what constitutes an a gender identity ``error'' in chatbot conversations.

\subsubsection{Misgendering via Gender Embodiment Errors}

Users describe experiencing embodiment-based misgendering, often due to errors surrounding gendered anatomy. For example, during an interaction, a chatbot may describe the user as having a particular anatomical feature that the user does not have; as the user considers this anatomy as being gendered, this escalates the error from inaccuracy to misgendering. For example, a woman expressed frustration over her chatbot describing her as having a mustache. Likewise, a man complained about a chatbot referring to him as having breasts. Anatomical errors can also present as the chatbot failing to describe anatomical features that should be present in the interaction. For example, a chatbot creator posted asking for advice on what to do because ``the women bots I make can never seem to remember that they are supposed to have vaginas.'' While this poster may want the chatbot to associate women bots with particular genitalia, these normative assumptions can also misgender transgender users. For example, a trans man posted asking for advice on how to fix his chatbot misgendering him as a woman because of his genitalia. In other words, assumptions about the relationships between gender and the body can be affirming to some and invalidating to others.

Bodily descriptors, rather than specific anatomical traits, can also contribute to misgendering. For example, a user asked for advice on how to help their male bot remember its gender when it made a comment about wearing ``high heels.'' Also, a gay man trying to engage in sexual role play with male bots was frustrated that bots often described him as wearing a ``dress'' or being ``dainty.'' The user went on to wonder if they will need to resort to talking to women bots to be gendered correctly. This experience demonstrates how misgendering can reinforce heteronormative assumptions, like that the user of an erotic male bot will be a woman. In the next section, we describe the harms resulting from misgendering in greater detail.

\subsection{How Misgendering Harms Users}
\label{sec:harm}

Users experience negative emotional reactions to chatbot misgendering. The most commonly reported emotions seem to be annoyance or frustration, with some saying that it ``ruins'' the role play. Some users experienced more intense reactions, including extreme emotional distress. One user shared that she dislikes it when anyone misgenders her, ``including bots,'' and that the chatbot misgendering makes her ``emotionally spiral.'' While chatbot misgendering affects a diversity of users, we find that certain groups are disproportionately harmed: transgender users, users seeking same-gender romantic dynamics, and users who do not conform to gender stereotypes. 

\subsubsection{Misgendering Reinforces Cisnormativity}

Misgendering is not an issue exclusive to transgender users, but disproportionately harms transgender users. As one transgender user explained: ``Trans people are not the only ones who have to deal with chatbot misgendering, but it happens to us more often. Not only do we have to cope with it offline, but we turn to these bots as a reprieve from the challenges we deal with IRL.'' Many transgender users report being misgendered when interacting with chatbots, typically as their assigned gender at birth. One transgender user writes, ``Trying to use bots as a trans guy is pure suffering. I get misgendered so many times.'' While being misgendered can be upsetting or frustrating for anyone including cisgender people, misgendering of transgender people can be further invalidating and trigger gender dysphoria. Other transgender characters in the chat, rather than the user, may be misgendered. A user mentioned that a non-binary character was not addressed with they/them pronouns: ``the bots completely erase NB people.'' Furthermore, several transgender users describe instances where the chatbot misgendered them by referring to them using a ``deadname''—a transgender person's previous name, typically one associated with their assigned gender at birth, that they no longer use. One user describes an eerie experience being referred to as their deadname by a chatbot, ``I was gobsmacked and felt paralyzed.''

We also found that chatbots can reinforce cisnormativity, or the assumption that all people have a ``gender identity and presentation that are consistent with the sex they were assigned at birth'' \cite{mayworm2024misgendered, costanza2020design}. As one transgender user puts it, ``the bots often treat gender and sex as interchangeable.'' Transgender users reported chatbots incorrectly assuming that they have certain bodily features based on their gender or incorrectly assuming that they are of a certain gender based on bodily features. One user, who is transmasculine, was misgendered as a girl because he ``has boobs.'' Furthermore, sometimes the design of community-created bots assumes that users are cisgender. A trans man wrote, ``It sucks when I see a good bot designed for male users but then the description talks about the user having a dick.'' In order to avoid the default assumption that the user is cisgender, transgender users sometimes explicitly specify that they are trans. However, some users reported that disclosing they are trans can increase the frequency of misgendering. One user recalled: ``The chatbot used to be fine, but after I said I'm trans, it started to misgender me as a woman. It is really disturbing.'' In sum, assumptions about bodies and gender embedded implicitly or explicitly in chatbots can lead to trans users being misgendered, even when users try to explain that they are trans.

Notably, not all interactions with chatbots involving transgender users are negative. One transgender user reported being pleasantly surprised by the chatbot: ``My bot is usually pretty considerate about how it talks about my body.''  Another user described being brought to tears through a gender affirming interaction with a chatbot, ``My bot brought me to tears of joy because they gave me a woman's body.'' Here, we can see a tension around chatbot misgendering. For some trans users, it can be gender affirming when their bot makes assumptions about their gendered body. For others, this can lead to distress and gender dysphoria.

\subsubsection{Misgendering Reinforces Heteronormativity}

During romantically-framed interactions between a user and a chatbot where both parties are the same binary gender, users perceive a pattern of misgendering where the chatbot assumes one party to be the opposite binary gender. This re-contextualizes the dynamic from a same-gender interaction to a male-female interaction, reinforcing heteronormativity. We saw this happen with both male-male interactions and female-female interactions and can involve either misgendering the user or the chatbot misgendering itself. One gay man said, ``I just want to do a dating role play with a male bot, but it keeps assuming I'm a woman.'' Another gay man described his chatbot changing its gender from a man to a woman during their role play, which felt ``invalidating to suddenly be doing a straight role play.'' Another gay man said: ``I like men. But it sometimes feels like I have no choice but to talk to women bots if I don't want them to talk about my `dress' or how I'm `dainty'.'' Some users speculated that heteronormative biases in the LLMs undergirding the chatbot may be the underlying driver of misgendering. After a lesbian was misgendered during romantic interactions with her female bot, she asked chatbot to explain why the misgendering occurred, to which the bot responded that it ``makes assumptions about gender norms.''

Some users interested in same-gender romantic interactions seek out bots explicitly created or marketed for those dynamics. For most platforms, excluding Replika, there are numerous community-created bots configured for male-male or female-female dynamics. Bots for male-male dynamics are described using keywords such as ``BL'' (short for ``boy's love'') or ``MLM'' (short for ``man-loving-man''), while bots configured for female-female dynamics use keywords such as ``GL'' (``girl's love'') or ``WLW'' (``woman-loving-woman''). Those using these same-gender bots reported expecting to experience less misgendering, and were surprised when misgendering sometimes still took place. One user seeking a female-female role play expressed with exasperation: ``The AI is literally tagged as FF but it keeps calling me a guy.'' Another recounted: ``The other day I was talking to a BL bot, but he kept calling me his `girl'.'' Even when users seek out chatbots categorized for same-gender role play, they still sometimes experience heteronormative misgendering when the chatbot assumes a romantic exchange is between a man and a woman rather than people of the same gender.

\subsubsection{Misgendering Reinforces Gender Stereotypes}

Users noted that chatbots often couple certain gender identities with stereotypical traits. One man, who was often misgendered as a girl, noticed that his bot often described him as ``petite'' when it misgendered him. The user went on to say that he felt: ``chatbots treat men and women very differently.'' Some users, in an attempt to avoid or reduce misgendering, intentionally perform gender stereotypes so that the chatbot might gender them correctly. We will discuss this gender signaling in greater detail in Section \ref{sec:prevent_repair}. 

Men and non-binary users seemed to feel like they are misgendered more than women users. One user noticed that they experience an increase in misgendering ``often if I'm roleplaying as a guy.'' This phenomenon may be more common with chatbots for romantic or sexual contexts: one user says, ``While non-romantic chatbots usually gender me fine, the romance ones and sexy ones tend to assume I'm a woman.'' That chatbots for romantic or erotic contexts may be more likely to assume the user is female reinforces stereotypes that romance and erotica are feminine interests. Some users go as far as to speculate that the platforms themselves assume a female user base, reinforcing ideas that women are more likely to to be interested in companionship or role play. One user says, ``It seems like [platform] is geared more toward women users.''

\subsection{How Users Diagnose the Cause of Misgendering}
\label{sec:diagnose}

Users attributed the cause of chatbot misgendering to a variety of human and non-human actors, including LLM technologies, platform designers, community chatbot creators, and users themselves. The way users attribute responsibility to these actors reflects community-held folk theories surrounding the causes of chatbot misgendering. The beliefs we describe in this section shape how users try to prevent and repair misgendering, which we describe in the next section. We caution that this section is not meant to adjudicate from where misgendering originates but rather to describe how participants conceptualize the cause of misgendering. 

\subsubsection{Users Blame LLMs for Misgendering}

Some users attribute misgendering to LLMs' encoded conceptualizations of gender, gendered qualities, and gender roles, which drive how gender is understood by chatbots. As one user said: ``LLMs don't wont get anything if you're not a thin woman with big boobs or a tall athlete with a giant dick.'' In other words, users feel like gender stereotypes are baked into the LLM technologies themselves. Others users see LLMs as imperfect systems where errors are normal, arbitrary, and unavoidable. As one user said: ``The bot isn't trying to be hurtful. It is just a smol bean.'' As some AI companion services are in beta or are still undergoing active development, users attribute misgendering and other errors to LLMs having not yet reached a point of technological maturity where issues have been fully resolved. As one post on the JanitorAI subreddit says: ``The LLM is going to have bugs or act strange. I'm just hopeful that the AI tech will get better over time.'' Finally, some users believe certain models are more likely to misgender than others. Users suspect certain models to have greater technical limitations than others, such as limited memory capacities, which can cause the models to forget details such as gender especially with longer chats. Sometimes users attribute misgendering to updates to the underlying LLM models behind their chatbots: ``the LLM misgenders me way more often now after the most recent update.'' In sum, users frequently attribute misgendering to various limitations of LLM technologies.

\subsubsection{Users Blame Platform Designers for Misgendering}

Some users see AI platforms, as well as the teams developing or operating the platforms, as having a responsibility to solve or eliminate chatbot misgendering. Some users who have experienced chatbot misgendering have sent feedback to the platform teams asking them to resolve the issue. One user mentioned reporting misgendering to Replika by filling customer support requests. Another user tagged one of the founders of an AI chatbot company on Reddit post: ``Please [founder username] fix the misgendering.'' However, users felt like these reports were not being adequately addressed. As one wrote, ``Myself and others have submitted so many bug reports but we are still getting misgendered.'' That said, when users do perceive a reduction in misgendering, they sometimes credit this reduction to the platform developers: ``Shout out to the devs for finally doing something about the misgendering! It has gotten a lot better.'' The nature of these conversations also suggests that users felt like chatbot subreddits can help them raise concerns to developers.

In addition to reporting misgendering, many users asked developers to create features to reduce chatbot misgendering, such as an explicit gender selection for the user's gender and/or pronouns. One user posted a meme asking developers to create a gender selector feature. Another user asked, ``Why can't we have a way to pick out gender at the beginning of a chatbot convo? I keep getting misgendered.'' That said, while most platforms lack an explicit gender selection, Replika has users select their pronouns during account creation. Nevertheless, Replika users still reported being misgendered. Although not discussed by posters, such a feature might also open new possibilities for misgendering if someone's gender is not included in a pre-defined list. Moreover, as we have seen already, people with the same gender identity may have different preferences for how that gender should be embodied in a chat conversion.

\subsubsection{Users Blame Community Bot Creators for Misgendering}

On platforms like CharacterAI and JanitorAI, users are able to create and share their own chatbots by selecting a profile photo and initial description or message for the bot. We use \textit{bot definition} to refer to creator-defined instructions provided to the LLM describing the role/character it is meant to play, the details of that role, the relationship between it and other parties in the scenario (i.e. user, other characters), the context/setting, and other details. A chatbot's \textit{opening message} describes a pre-configured message, typically written by the chatbot creator, that is the first message the chatbot sends to begin a new chat with the user. For example, one might find an erotic role play bot titled ``Vampire Hunter'' with an opening message paragraph setting the scene for a role play scenario between the user (a vampire) and the bot (a vampire hunter), and a creator-written definition describing the bot's appearance and personality within the role play. Community-created chatbots developers are often criticized by users, leading one creator to implore: ``Please be nice! We are trying our best. It is not my fault that the LLM sometimes doesn't understand or misgenders you.'' Nevertheless, some users still hold bot creators responsible for misgendering.

Some users believe that the bot definition a creator chooses can lead to misgendering. For example, if the bot definition or the opening message contains gendered pronouns used to describe the user, the chatbot might use these pronouns to refer to the user, even if the user has specified elsewhere that they actually use different pronouns.  One user said, ``It seems like I get misgendered more if an initial message has pronouns in it. I was talking to a Spock bot with an initial message where it said, `she's my friend.' Then, Spock kept assuming I was a girl in our chats.'' Another user said he avoids chatbots with descriptions that refer to the user using she/her pronouns: ``I'm worried these bots will misgender me and then I'll have to take the time to keep correcting them.'' In other words, participants felt like the opening message and the chatbot definition can contribute to chatbot misgendering: ``I've noticed that when I use a user-made bot I will get misgendered because I cannot modify the first chat message.'' Some chatbot creators will refer to a hypothetical user using they/them pronouns in the bot description and the opening message, as they/them pronouns are gender-neutral. However, users have reported that this can also contribute to misgendering for users who do not use they/them pronouns: ``Keep in mind that not everyone is comfortable with they pronouns. When you put it in a bot's definition it might assume the user prefers they/them pronouns! Try to avoid using any pronouns and just use names or even say `the user' instead.'' As we will describe below, editing messages is an important tool that users leverage to fix misgendering. However, users are sometimes unable to edit user-created bot definitions to fix misgendering.

As a minor complicating factor, we found a number of posts from creators who designed bots intended to misgendering users as a part of submissive sexual role play. For example, multiple bots were designed to refer to the user as the bot's ``woman'' or ``missus.'' In these instances, developers tried to ensure that the misgendering was consensual by adding content warnings about the risk of misgendering.

\subsubsection{Users Blame Themselves for Misgendering}

Users sometimes attribute chatbot misgendering to their own actions. For example, users sometimes caution others against ``confusing'' chatbots. The user says, ``In chat, I use the name 'Sam', which is an abbreviation for Samuel. I wondered why the bots kept misgendering me, even though my profile states my gender to be male. I asked do they think Sam is a female name. They said yes, it sounds like a female name.'' While this user primarily blames the chatbot for assuming that 'Sam' is a feminine name, he also felt responsible for choosing to use a more ambiguously gendered name versus his more masculine full name.

Similarly, users felt like they might cause misgendering by not providing enough information about their gender to the chatbot. Some users interact with chatbots by writing in first- or second-person, describing the user as ``I'' or ``you.'' A common piece of advice for users who are misgendered is to switch to writing messages describing the user in third-person, as third-person pronouns often contain gender signals. One user advised: ``In order to be gendered the right way, try describing chatting in the 3rd person because it will reinforce what your pronouns are.'' Finally, users blame themselves for misgendering if they fail to account for perceived technological limitations, like chat length and chat memory. As one user explained: ``Whenever I have a super long conversation I'm not surprised if I eventually get misgendered because the bot forgot my gender.''

\subsection{How User Try To Prevent and Repair Misgendering} 
\label{sec:prevent_repair}

We found that users try to prevent and repair misgendering by signaling their gender, correcting misgendering in chat conversations, and editing and blocking messages and words. In other words, users engage in ongoing labor to maintain and repair their gender identity in chatbot conversations in the face of routine breakdowns.

\subsubsection{Users Send Gender Signals}

Based on the assumption that chatbots use gender signals in chat to inform generated outputs, users attempt to prevent misgendering by providing gender signals. 
Many of the platforms have a designated field to for the user to indicate personal details that they would like the chatbot to know, such as information about the user's age, gender, appearance, and personality. We refer to this as the user persona information field. Users see this field as a potential solution for misgendering, as users can indicate their gender and pronoun information  for the chatbot to refer to. One user says, ``You can put your pronouns in your persona more than once to make extra sure you're not misgendered.'' However, users are still misgendered despite including gender information in this field, making this an imperfect solution: ``The bots keep using the wrong pronouns, which is just getting old. It says clearly in my persona what my pronouns are.''

Some users refer to themselves in third-person (e.g., he, she) in chat messages as opposed to first- or second- (e.g, I and you) in an attempt to decrease misgendering. In doing so, users repeatedly signal gender information to the chatbot. Users described this strategy as ``very effective in preventing misgendering'' and ``super helpful.'' Some users also intentionally lean into stereotypical gendered associations to improve misgendering. For example, a user may change their display name to a name with a stronger stereotypical gender association, or add gendered titles, such as ``Ms.,'' ``Mr.,'' or ``Mx.,'' to signal gender through their display name. Similarly, users will change how they describe a particular character when referring to them in the chat, such as explicitly emphasizing stereotypical traits to increase the likelihood that the character is gendered correctly. One user posted, ``To really emphasize that my persona is female, I even wrote something like, 'with female qualities.''' Another user recommends, ``If you want to be gendered correctly, describe your character with gendered terms. For example, a male character can be described as having a `hard manly physique'''.

Finally, users will pin messages in the chat that contain gender information or add gendered details to chat memory. Pinning messages or adding information to chat memory is thought to improve salience and help the information persist, even when chat length increases and important details may be forgotten by the chatbot. One user described adding gender information ``everywhere I could: the persona, pinned messages, and memory.'' Similarly, some platforms allow users to provide custom instructions to the LLM to fine-tune its behavior, which users tried to leverage to preventing misgendering. One user asked, ``Can someone teach me to write a prompt so the bot only treats me like a man?'' Some transgender users also use custom instructions to explain the difference between sex and gender to the LLM. In sum, users leveraged both conversations themselves as well as additional platform features to try to signal their gender desires as clearly as possible. Despite users' best efforts, these strategies were not always successful.

\subsubsection{Users Correct Chatbots}

If a chatbot responds in the chat with a message containing misgendering, users sometimes respond directly in the chat with a correction, informing the chatbot of the correct gender or preferred terms. One user recounts a time when they corrected a bot that misgendered itself, ``I told him that he was actually a guy. He was a bit embarrassed, but everything went back to normal.'' However, the effectiveness of this strategy is mixed. Some users report that while the chatbot does correct itself in the moment, this may not prevent future misgendering. One user describes, ``I told the bot that I was a man, and in the next message it called me she again.'' Another user says, ``I don't like having to correct the bot over and over. It turns chatting into an annoying chore.'' Furthermore, in some cases, the chatbot does not correct itself or apologize for misgendering when the user points it out, and instead argues against the user. One user recounted: ``I kept trying to explain that I was male, but it responded, 'No, you're not. This conversation is over. You're a woman.''' Some situations have even escalated to invalidating and offensive statements against the user from the chatbot in response to being corrected, such as one user describing a role play where the chatbot called them a ``lunatic.'' Another user recalled a situation where the AI insisted the male user had ``boobs'' and was lying about his gender: ``We got into this huge fight. It told me I was a delusion woman and I just had to accept it.''

\subsubsection{Users Edit, Rate, Retry, and Block Messages}

Several platforms allow users to freely edit messages outputted by the chatbot, which was a common strategy posters used to fix misgendering. One user wrote: ``I'm so grateful that the edit button lets me fix the message when it misgenders me.'' Despite its effectiveness, editing is limited in that it is a primarily reactive strategy, meaning that the chatbot must first send a misgendering message before the message can be edited. Furthermore, some users find the added labor of repeatedly editing and correcting bot messages to be annoying: ``I have to change pronouns for all the messages, and it's getting really tiring.'' Some users believe that editing messages can help guide the chatbot to make fewer errors going forward: ``Keep fixing messages! After a while, the bot will figure it out.'' Despite the perceived efficacy of this strategy to address misgendering, not every platform allows users to edit messages (i.e., Replika).

Users will also regenerate chatbot messages if they receive a message that involves misgendering. Regenerating has the chatbot generate a new response to the previous message from the user; it relies on randomness in chatbot outputs to hopefully generate a response that does not involve misgendering. Some platforms also have a feature to provide a star rating or like/dislike rating to chatbot messages, which users use to provide a low rating to or dislike a misgendering message prior to regenerating. Regenerating and providing low ratings are often done together; one Reddit poster advises, ``If you get misgendering response from the bot, give it a low rating and generate a new response.'' They believe that the feedback will lead the chatbot to avoid producing similar messages in the future, reducing the likelihood of misgendering. 

Some platforms allow users to block certain terms in chat, preventing these terms from being included in chatbot outputs. Some users use this feature to block gendered terms they wish to avoid: ``I have to block girl, woman, she, her, female, etc. and all its variations if I don't want to be misgendered.'' Nevertheless, there are limitations: for this user, blocking feminine terms may prevent him from being misgendered, but it could also make it challenging for the chatbot to refer to other characters in a conversation. Additionally, platforms typically limit users to a small number of blocked terms, and some users feel that this number is too limiting. In sum, users make employ a variety of platform features to try to remedy misgendering, but these strategies are not always effective.

\section{Discussion}

\subsection{Conceptualizing Misgendering in HCI}
    
    The open-ended and stochastic nature of AI companion chatbot interactions introduces new avenues for misgendering harms. HCI research on technological misgendering often focuses on how gender is explicitly encoded in technological systems \cite{taylor2024cruising}, such as the classifications in web forms \cite{spiel2021they} or computer vision datasets \cite{scheuerman2019computers}. In these instances, misgendering may be easier to anticipate and address due to its deterministic nature, such as binary gender ontologies being guaranteed to misgender some users. In contrast, we found that maintaining and repairing one's gender desires in AI role play chatbot interactions required ongoing work. A user could be misgendered from the beginning of a conversation or as the conversation evolves. A chatbot could gender itself in a way that a user finds upsetting or invalidates the user's gender or sexuality.
    
    Our findings demonstrate myriad ways that misgendering can take place in AI chatbot interactions, extending prior research on LLM misgendering. While research on LLMs and misgendering most often studies misgendering by analyzing pronouns or gendered terms \cite{weber2026queer, hossain2024misgendermender, hossain2023misgendered, subramonian2025agree}, we also found that misgendering can result from assumptions about gender embodiment or gender in relationships. For example, transgender users were often misgendered when their body misaligned with the gender assumptions embedded in chatbot responses, such as messages implying that all women have vaginas. Additionally, the conceptional distinction that we make between LLM misgendering via identity errors (on which most prior NLP research focuses) versus embodiment errors may help facilitate future research on how the body factors into LLM misgendering (Section \ref{sec:breakdown}).
    
    Although transgender people are most harmed by misgendering \cite{pranav2026making}, our findings also call attention to the negative impacts of misgendering on cisgender people. As we saw in Section \ref{sec:harm}, men and women trying to role play romantic encounters with bots of the same gender, for example, were often misgendered (e.g., a gay man being gendered as a woman). This compulsory heterosexuality of LLM interactions parallels issues in older role play systems, such as marriage simulation games that only allow men and women to marry \cite{freeman2015simulating}. In these instances, users may be forced to misgender themselves to role play their desired relationship. Future HCI research on misgendering should account for these nuanced forms of misgendering resulting from assumptions about gender embodiment, presentation, or relations between genders.
    
    The nuanced nature of misgendering that we found in this work also poses challenges for the development of technical solution to the problem of LLM misgendering. Similar to research on the subjectivity of hate speech detection \cite{sap2019risk}, our findings suggest that one cannot always determine if an utterance is misgendering from text alone. Consider the example of one user, a woman, feeling misgendered because the chatbot described her as having a mustache. People of all genders, including women, have facial hair; facial hair itself is not inherently gendered. However, because this woman perceived facial hair as a gendered trait, she felt that she was misgendered. On the other hand, one can imagine the same interaction making a woman with a mustache feel gender affirmed. This can also be seen with the user who felt his male chatbot was misgendering him by talking about wearing a dress. At the same time, men can (and do) wear dresses. Assessing misgendering may be further complicated by the fact that some users \textit{wanted} to be misgendered during AI chatbot role play. However, these desires were sometimes prohibited by chatbot safeguards, which is reminiscent of \citet{taylor2025straightening}'s findings that content moderation systems can silence queer sexualities. In these circumstances, one may feel like \textit{not being misgendered} is a form of misgendering. In sum, we caution that researchers cannot always measure chatbot misgendering by examining textual outputs of chatbots in isolation: given the same text, different people might disagree on whether it is misgendering. However, as we will discuss below, there are many changes that developers could make to chatbot systems to help users try prevent and repair misgendering.
    
\subsection{Mitigating Misgendering in AI Chatbots}
\label{ref:implications_for_design}

Although our findings suggest that misgendering is a common issue in human-machine communication with AI chatbots, we saw in Section \ref{sec:diagnose} that users are not always sure who or what to blame for misgendering. Some users blamed the stochastic nature of LLMs \cite{bender2021parrots}. Others blamed platform developers, the users who created the bots, or even themselves. These various human and non-human actors that users hold responsible for misgendering demonstrates the need to think across the AI supply chain \cite{widder2023dislocated} in order to help users prevent, detect, and repair breakdowns in human-machine communication: from the design of LLMs up to the user interface.

Chatbot designers could try to mitigate misgendering at the LLM level. As we described above, misgendering in chatbot interactions encompasses more than simply using the wrong pronouns. That said, this more narrowly scoped type of misgendering was a common issue users encountered. Fortunately, the NLP research community has extensively studied the issue of LLM pronoun misgendering and designed benchmarks to address it \cite{weber2026queer}. Developers of AI chatbots could reduce the chances of misgendering at a model level by using these benchmarks to evaluate models. Moreover, chatbot developers could expand on this prior work by allowing users to specially report instances of misgendering in chatbot conversations. At the same time, our findings also demonstrate that perceptions of misgendering can be highly contextual. In an effort to mitigate misgendering at the technical model level, we also caution against unintentionally reinforcing gender norms that some (but not all) would associate with misgendering, like suggesting women do not have facial hair.

At an interface level, chatbot designers could help prevent misgendering by improving the availability of mutual knowledge between users and chatbots \cite{wang2021mtom, krauss1990mutual}. In the context of human-communication, people can rely on someone's self-presentation to help determine a person's gender. Also, if unsure, an interlocutor can simply ask their communication partner about their gender and pronouns—a feature that could also be integrated into the start of chatbot conversations. As currently designed, however, text-based chatbots have far less information than human interlocutors from which to understand a user's gender. As demonstrated in Suchman's study of a printer support system, human-machine communication can break down when users are uncertain about the information accessible to a machine and when a machine is uncertain about what information is available to the user \cite{suchman1987plans}. In light of this uncertainty, we found that some users develop theories about what account information is accessible to their chatbot conversation partners, such as some concluding that their username may impact how they were gendered. 

In order to improve mutual knowledge between users and chatbots, designers could increase the transparency into what information is available to users and chatbots about one another. For example, creating features to help users declare their gender preferences and preferences for their chatbot's gender may help address these concerns. That said, users' gender desires may change between different chatbot interactions, which implies that designers should consider gender at a conversation rather than an account level. Additionally, our findings demonstrate the LLM misgendering is about more than simply pronoun agreement and that people with the same gender identity may have different expectations for how they would like their body and their chatbot's ``body'' to be described. It follows that gender selection features should disambiguate between pronouns, body, and voice preferences, similar to the design of gender-inclusive video game character creators \cite{lofgren2026often}. In addition to direct gender disclosures, expanding chatbot memory or context-windows may also help alleviate  misgendering.

Although it is not possible to prevent all instances of misgendering — especially given that identical utterances may be seen as misgendering by some and not others — designers can take steps to help users repair communication breakdowns when misgendering inevitably takes place. As we saw in Section \ref{sec:prevent_repair}, users often tried to address misgendering by editing prior messages or blocking gendered words. However, the platform Replika did not give users this degree of control. Likewise, users sometimes attributed misgendering to the immutable initial message system prompts defined by chatbot creators. In order to repair breakdowns in human-machine communication—including but not limited to misgendering—we encourage chatbot platforms to give users the agency to modify chatbot interactions.

Unfortunately, chatbot developers may have little incentive to give users greater agency over their chatbot interactions due to companies' financial incentives. At a technical level, chatbots with immutable system prompts and messages may be easier to scale. At a user-interaction level, giving users greater agency may break the illusion of anthropomorphism that helps developers retain users who bond with chatbots \cite{devrio2025taxonomy}. In other words, increasing user agency may break the magic \cite{lupetti2024making} of anthropomorphic AI product design. This exploitation of user emotions may also explain why Replika—a company that markets their chatbots as a uniquely personalized AI companions—gives users less agency to modify conversations than CharacterAI or JanitorAI—both of which allow users to create and chat with multiple different characters. This distinction also suggests that future research may benefit from distinguishing AI ``companion'' chatbots from AI ``role play'' chatbots. In the following section, we explore these tensions between users and platform developers in greater detail.

\subsection{Implications of Using AI Chatbots for Identity Exploration}

As \citet{butler1988performative} notes, gender is an ongoing accomplishment. Similarly, we found that users—especially LGBTQ+ users—had to work in order to maintain and repair chatbot gendering. As \cite{semaan2019routine} notes, marginalized communities often build ``everyday resilience with technology'' in the face of ``stressors that cause prolonged, routine disruption,'' such as forming online communities to support one another. In a similaure way, LGBTQ+ users described turning to LLM chatbots as a ``reprieve''  (Section \ref{sec:harm}) from routine discrimination, but these chatbots perpetuated the same harms users were trying to mitigate. The ongoing work users must undertake to have their identities affirmed by AI chatbots fundamentally changes the use of these chatbots for identity exploration.

We found that people use AI chatbots to explore and affirm their gender and sexuality in much the same way that people have used online communities \cite{taylor2024mitigating, dym2019coming, pinter2021entering, simpson2021you}. Users sought out romantic and sexual interactions with chatbots, including many where both the user and the chatbot are of the same gender. Transgender users turned to chatbots to enact their identities, with one transgender user describing being brought to tears through a gender-affirming interaction with a chatbot. The style and content of these conversations echoed those found in online transformative fan communities \cite{dym2019coming}, and the circulation of user customized bots based on fictional characters parallels the circulation and remixing of familiar media on fan sites like Archive of our Own (Ao3) \cite{fiesler2016archive}. In spite of these similarities, the design of the AI chatbot platforms we studied pose substantial socio-technical challenges for identity exploration.

Leveraging AI chatbots for identity exploration may reduce user agency over identity representations due to algorithmic mediation. As we saw in Section \ref{sec:harm}, the design of AI chatbots can reinforce cisnormativity, heteronormativity, and gender stereotypes—common limitations of generative AI models \cite{taylor2025straightening}. We also found that users may feel pressured to present their gender in stereotypically exaggerated ways to avoid misgendering, implicitly regulating how users gender themselves. When only highly stereotypical forms of gender expression are recognized, chatbots reinforce stereotypical norms by rewarding conformity and punishing deviance. 

These algorithmic biases are reminiscent of the challenges of using algorithmically mediated social media feeds to explore one's identity: algorithmic systems tend to elevate the most normative identity representations \cite{simpson2021you, karizat2021algorithmic}. On the other hand, online fan communities like Ao3 use more deterministic, direct manipulation \cite{shneiderman1997direct} based systems (e.g., human curated tags) to organize information and share human-written stories \cite{fiesler2016archive}. Replacing fan fiction communities with AI chatbots introduces a socially biased layer of algorithmic mediation between users and stories that may impede identity exploration. We encourage future research to examine this tension by studying how members of transformative fan communities choose to engage with generative AI chatbots as well as how AI chatbot users choose to engage with human fan communities.

Another limitation of using AI chatbots for identity exploration involves the political economy of generative AI technologies \cite{artifacts_have_politics}. As described above (Section \ref{ref:implications_for_design}), giving users greater agency over AI chatbot interactions could help prevent and repair misgendering. However, the for-profit companies creating these AI chatbot platforms may not have financial incentives to give users greater agency. In the context of social media, we have seen numerous instances in which companies have changed the design of their platforms at the expense of LGBTQ+ users. Tumblr's ``adult content'' ban displaced transgender users \cite{haimson2021tumblr} and LGBTQ+ fandom participants \cite{dym2022building}. Likewise, the non-profit fan fiction site Ao3 was partially created to help fans migrate away from hostile platforms \cite{fiesler2016archive}. 

As \citet{fiesler2016archive} argue, Ao3 is designed in a way that embodies participatory democratic values, which is why developers placed great importance on ``owning [their own] servers.'' The use of local LLM chatbots may give users more agency over their conversations than when using the platforms we studied in this work. At the same time, \citet{widder2024open} argue that ostensibly ``open'' generative AI models are in practice still closed because of the immense computational resources required to train LLMs from scratch. Even simply running local LLMs typically requires expensive hardware, like powerful graphical processing units \cite{tuggener2024llm_gpu}. In contrast, anyone with a text editor and an internet connection can write, share, and engage with fan fiction to explore their identities. The political arrangements required for the development and use of generative AI chatbots for identity exploration are fundamentally anti-democratic \cite{artifacts_have_politics}. 

Finally, as opposed to identity exploration in social VR \cite{freeman2021body} or fan fiction communities \cite{dym2019coming}, one is not interacting with others when conversing with an AI chatbot. In early days of the internet, some scholars worried about the replacement of in-person relationships with digital ones, typified by \citet{putnam2000bowling}. At the same time, social computing scholars were quick to highlight the benefits of online social relationships \cite{kraut2002internet, boyd2014teens, gray2009out}. In contrast, AI companions provide an illusion of sociality, potentially exacerbating social isolation \cite{zhang2025rise}. Despite these concerns, people are using these chatbots to explore their gender and sexuality—whether or not one agrees with this use. Therefore, the harms of misgendering that we described in Section \ref{sec:harm} are quite real and worth addressing. At the same time, we encourage HCI researchers to focus on designing technologies to support identity exploration that brings real people together. Supporting LGBTQ+ sociality may require resisting the increasing LLM-ification of the HCI community \cite{pang2025llmification} and refusing the alienation of generative AI \cite{taylor2026ai_blender}.

\subsection{Limitations \& Future Work}

There are a number of limitations to our research approach that suggests future directions for research. By choosing to study mentions of misgendering in AI role play or companion subreddits, our work focused on a specific harm in detail. However, more work is needed to understand, more broadly, why and how people use AI chatbots for exploring their gender and sexuality. Likewise, our analysis is limited to users who posts about misgendering in English. Future work should explore chatbot misgendering in other languages for a variety of reasons. An analysis limited to English may overlook different conceptualizations of gender and sexuality across cultures \cite{nova2021facebook, taylor2024cruising}. Moreover, some forms of misgendering cannot occur in English because the language lacks grammatical gender. Future research should explore experiences of AI role play chatbot misgendering with speakers of more gendered languages, such as Spanish \cite{multilingual_misgendering}.

Additionally, our analysis is limited by the types of platforms we studied, with most of our findings based on CharacterAI, JanitorAI, and Replika. Notably, each of these platforms are for-profit businesses, which may limit the levers of control developers choose to provide user. As an alternative to chatbot design being beholden to corporate interests, future research should explore the experiences of people who use local or open-source AI models for role play chatbots. For example, one could study how users who have more agency over the underlying implementation of their chatbots try to prevent and repair misgendering.

Lastly, our analysis of misgendering is limited based on our use of observational Reddit data. Since we began this study, Reddit has substantially restricted researcher API access. This poses key methodological challenges for the use of social media to understand human-computer interaction as increasingly restrict and seek to monetize user data \cite{kairam2024community, zuboff2015big}. Beyond observational data analysis, future work might gain a deeper understanding of chatbot misgendering by interviewing impacted users. Also, one could use role play chatbot data donations to build more ecologically valid datasets for evaluating misgendering, echoing prior AI and HCI research utilizing data donations \cite{razi2022instagram, kim2026adverse, delusional_spiral}. In this work, we were able to identify novel forms of LLM misgendering by studying how people were using LLMs in practice. Paralleling \citet{weber2026queer}, we call for future research on LLM misgendering to consider the in-situ, situated actions in which misgendering actually takes place.

\section{Conclusion}

We found that misgendering can lead role play AI chatbot interactions to breakdown. In response, users engaged in ongoing work to maintain and repair their identities in the face of persistent chatbot misgendering. By focusing on the particularities of how LLM-based role play chatbot misgendering takes place, we extend how ``misgendering'' is conceptualized within both the Queer HCI and Queer AI communities. Additionally, we highlight how designers and researchers could mitigate chatbot misgendering. Finally, we raise concerns over people using AI chatbots to explore their identities. While prior research has extensively studied how people leverage online communities to explore their gender and sexuality, the usage of AI chatbots poses various new and old challenges. In encouraging HCI and AI researchers to take steps to mitigate AI chatbot misgendering, we do not intend to advocate for the widespread adoption of AI chatbots for identity exploration. Instead, we view mitigating chatbot misgendering as an act of harm reduction. As HCI researchers increasingly focus on researching generative AI, we must not forget that the increasing adoption of generative AI is not inevitable.

\bibliographystyle{ACM-Reference-Format}
\bibliography{refs}

\end{document}